\documentclass[final,conference,twocolumn,letterpaper]{IEEEtran}

\IEEEoverridecommandlockouts              

\usepackage{fancyhdr}
\usepackage{graphicx}
\usepackage[utf8]{inputenc}

\graphicspath{{figures/}}

\usepackage{url}
\usepackage{wrapfig}
\usepackage{units}
\usepackage{caption}
\usepackage[printonlyused]{acronym}

\usepackage{calc}
\usepackage{graphicx}
\usepackage[lined,boxed, vlined,ruled,linesnumbered]{algorithm2e}
\usepackage{amsmath,amssymb,amsthm,amsfonts,amsbsy}

\usepackage{algpseudocode}
\usepackage{color}
\usepackage[table]{xcolor} 
\usepackage[caption=false]{subfig}
\usepackage{dblfloatfix}
\usepackage{nicefrac}
\usepackage{tcolorbox}
\usepackage{txfonts}

\usepackage{multirow}

\newcolumntype{C}[1]{>{\centering\arraybackslash}p{#1}}

\usepackage{booktabs}

\definecolor{BgGray}{gray}{0.7}%
\definecolor{BgGray2}{gray}{0.96}%
\definecolor{RowColorOdd}{named}{BgGray2}%
\definecolor{RowColorEven}{named}{white}%
\definecolor{comments}{gray}{.5}
\definecolor{Gray}{gray}{0.85}

\providecommand{\keywords}[1]{\textbf{\textit{Index terms---}} #1}

\usepackage{pifont}
\usepackage[multiuser,nomargin,marginclue,footnote]{fixme}
\fxusetheme{color}
\fxsetup{targetlayout=colorcb}
\FXRegisterAuthor{all}{anall}{ALL}
\FXRegisterAuthor{md}{anmd}{MD}
\FXRegisterAuthor{tolja}{antolja}{TOLJA}
\FXRegisterAuthor{pg}{anpg}{PG}
\FXRegisterAuthor{mch}{anmc}{MC}
\FXRegisterAuthor{cp}{ancp}{CP}
\FXRegisterAuthor{awo}{anawo}{AWO}

\fxsetup{status=draft}

\begin{document}
\providetoggle{techreport}
\settoggle{techreport}{false}

\title{WoV: WiFi-based VLC testbed}

\author{
\IEEEauthorblockN{Piotr Gawłowicz, Elnaz Alizadeh Jarchlo, Anatolij Zubow}
\IEEEauthorblockA{\{gawlowicz, jarchlo, zubow\}@tkn.tu-berlin.de}
Technische Universität Berlin, Germany\\

}

\maketitle

\begin{abstract}
We present a complete Visible Light Communications (VLC) trans\-ceiver system consisting of low-cost Commercial-Off-The-Shelf (COTS) components.
In particular, we show that COTS IEEE 802.11n (WiFi) devices can be used so that the physical and data link layers of radio frequency (RF) WiFi, i.e. 2.4\,GHz, are reused for VLC.
Moreover, as WiFi is fully integrated with the Linux system, higher protocols from network to transport and application layer can be used and tested in VLC-related experiments.
Our approach has the advantage that a VLC experimenter can fully focus on VLC-related low-level aspects like the design of novel VLC front-ends, e.g. LED drivers, lenses, and photodetectors and test their impact directly on the full network protocol stack in an end-to-end manner with real applications like adaptive video streaming.
We present first results from experiments using our prototype showing the performance of unidirectional VLC transmission.
Here we analyze the distortions introduced as well as the relationship between signal strength on frame error rate for different MCS and the maximum communication distance.
Experimental results reveal that a data rate of up-to 150\,Mbps is possible over short ranges.
\end{abstract}


%
\section{Introduction}
To satisfy the exponentially increasing demand for mobile data communications radical new solutions are needed.
One promising idea is to off-load some of the data traffic from the radio frequency (RF) domain to the optical domain, e.g. Visible Light Communications (VLC), and use the precious RF spectrum in applications most needed.
With the wide spread use of light emitting diodes (LEDs) in smart devices, flashlight, street and traffic lights, vehicles, trains, planes, etc. there is the opportunity to set-up VLC links for a range of applications in both indoor and outdoor environments. 
However, VLC is very sensitive to the blockage of objects and suffering shadowing due to the high directionality of the optical channel.
Hence, VLC has gained substantial attention from both industry and the research community which require large-scale deployments of low-cost VLC solutions indoors (apartments, industry buildings) as well as outdoors.

In this paper, we present WoV (\textbf{W}iFi-\textbf{o}ver-\textbf{V}LC), a complete VLC testbed consisting of Commercial-Off-The-Shelf (COTS) components.
We show that unmodified COTS 802.11n WiFi devices designed for operation in RF can be used to conduct VLC experiments and hence allowing the experimenter to focus on low-level VLC designs like LED, driver circuits, lenses, photodetectors.
As the proposed hardware is cheap large-scale VLC testbeds can be built to analyze the performance of VLC on both link (e.g., antenna diversity techniques) as well as system-level (e.g., handover and performance in interference channel).
Moreover, as capturing of channel state information (CSI) is possible with WiFi such information can be used to study channel characteristics of VLC.
Finally, as a full network stack (from PHY to transport) is available end-to-end manner experiments with real applications (like \texttt{iperf}, video streaming, etc.) are possible.
Finally, our work is in line with the goals of the upcoming IEEE 802.11 Light Communication (LC) standard, where the objective is to amend the Medium Access Control (MAC) and Physical Layer (PHY) of IEEE 802.11 with Light Communications~\cite{80211bb}.
\smallskip

\noindent\textbf{Contributions:}
We present a simple and inexpensive COTS-based evaluation platform for prototyping VLC front-ends.
Specifically, we exploit the vast set of capabilities already implemented in modern RF WiFi chipsets.
Therefore, a researcher can focus entirely on the design of novel VLC front-end transceivers.
We provide proof of concept prototype implementation and perform evaluation in a small testbed.
Using our prototype we were able to quickly investigate the impact of changes made to the VLC-frontends on end-to-end performance, e.g. introduction of simple optics (lenses) on the transmitter side greatly extends communication distance.

%
\section{Background}
In this section, we give a brief introduction to VLC \& WiFi.
\subsection{Primer on VLC}
Visible Light Communications (VLC) which also is known as Li-Fi~\cite{arnon2015guest}, is a short-range wireless communication technology that is gaining momentum in research and industry for both indoor and outdoor network environments. 
VLC is considered as a complementary technology to radio-frequency (RF) communications and it is well suited for higher link data rates and it promises a reliable connectivity due to its very large spectrum between 375 and 780 nm.
VLC uses light emitting diodes (LEDs) as transmitters which illuminate with the sufficiently high rate of modulation and the flickering is undetectable by the human eyes. 
The intensity of the emitted light is modulated with the desired information using various modulation schemes such as on-off keying (OOK)~\cite{lee2011indoor,miramirkhani2015channel} and/or orthogonal frequency-division multiplexing (OFDM) and finally is received by photodiodes (PDs). 
Note that the input signals have to be non-negative unipolar with the real values, as the light intensity cannot be negative.
VLC has a numerous advantages as it does not have interference with RF cellular networks and it has an increased capacity thanks to its large spectrum.
Moreover, VLC provides a high bandwidth which implies good spatial resolution e.g. for wireless positioning.
In addition, in terms of energy and cost it is efficient and most importantly it is a high secure technology since it does not penetrate through walls~\cite{sarbazi2014ray}.
As LEDs in the buildings, vehicles, and consumer products are rapidly gaining visible light communication capabilities, the main challenges for VLC can be connectivity disruption due to shadowing, blockage, mobility and, the external light~\cite{wang2017light}. 
Following, the loss of communication causes a frequent handover in the network. 
Additionally, communication in large range and outdoor scenario would be one of the issues for VLC, since it leads to several problems such as hidden node and chance of collision in the network~\cite{matus2017hardware}.

\subsection{Primer on WiFi}
An essential characteristic of the IEEE 802.11 specification is that there is a single medium access control (MAC) sub-layer common to all physical (PHY) layers.
This feature will allow easier interoperability among the many physical layers.
There are already plenty of physical layers in the standard: infrared, frequency hopping spread-spectrum, direct sequence spread-spectrum (DSSS) and orthogonal frequency division multiplex (OFDM).
Infrared never took off especially because of its low data rate, etc.
However, infrared and radio can be considered as complementary technologies for the support of WLANs. 
We believe that VLC can take over the role of infrared technology as it is well suited for low-cost low-range applications, such as ad-hoc networks.

The most widely used 802.11 physical layer is based on OFDM.
The default channel has a bandwidth of 20\,MHz.
Moreover, there are options to aggregate adjacent channels together, i.e. up to 2 and 8 channels in 802.11n and 802.11ac respectively.
To account for frequency-selective channels WiFi performs equalization in frequency domain.
The total number of OFDM subcarriers depends on the WiFi standard and number of aggregated channels.
It can range from 64 to 256 for a 20\,MHz channel in case of 802.11n/ac and 802.11ax respectively.
WiFi transmits data as self-contained asynchronous frames which can be independently detected and decoded thanks to the prepended preamble and PLCP header (i.e. control data), respectively.
The WiFi frame duration is bound, e.g. to 5.484\,ms in 802.11n.
The TX power can be set on a per frame basis which is possible with most COTS WiFi chips.
For multiple access stations perform random channel access using a Listen-Before-Talk (LBT) scheme (i.e. modified CSMA).
Coexistence among multiple WiFi sets is achieved using both virtual (reservation) and physical carrier sensing.
The former is used to combat well-known problems like hidden terminal where the channel is reserved before usage by means of exchanging small signaling packets reserving the channel, i.e. RTS/CTS.

WiFi is a mature and well-established technology.
Hence COTS WiFi chips offer several additional functions.
With most WiFi NICs it is possible to measure high level statistics like frame error rate, receive signal strength (RSSI), noise floor.
Moreover, some chipsets (e.g. Intel IWL5300) provide channel state information (CSI) measurement capabilities which are of great importance for WiFi-based localization as they provide detailed information on the effect of the radio channel.
Other WiFi NICs like the one from Atheros provide spectral scan capabilities which can be used to estimate the receive power on the level of OFDM subcarrier.
%

%
\section{WiFi over VLC} \label{sec:arch}

In this section, we first present the WoV architecture and describe the hardware devices and the software configuration used.
Afterwards selected experimental results obtained from the proposed WoV testbed are presented.
\subsection{VLC Architecture}

\begin{figure}[ht]
    \centering
    \includegraphics[width=0.95\linewidth]{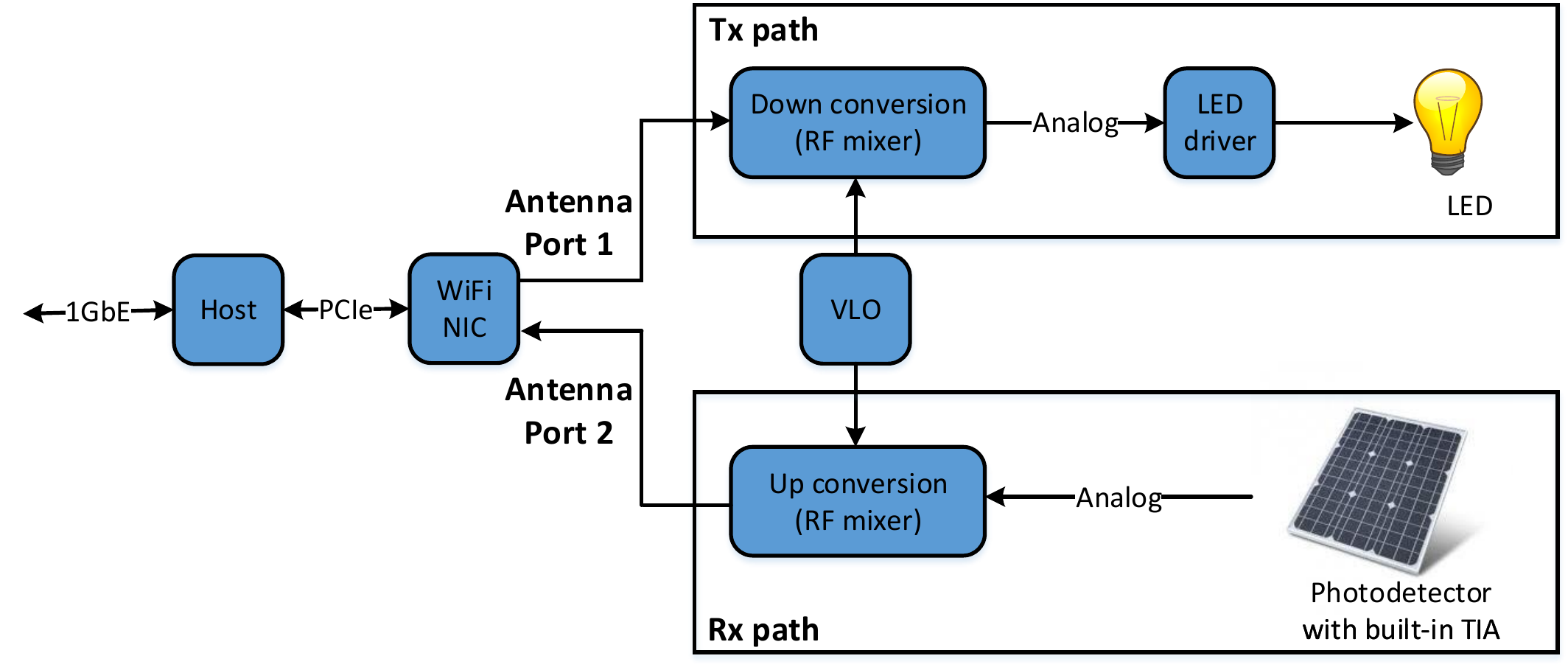}
    \vspace{-5pt}
    \caption{Architecture of WiFi over VLC.}
    \label{fig:wifi_over_vlc_arch}
    \vspace{-5pt}
\end{figure}

Fig.~\ref{fig:wifi_over_vlc_arch} shows the schematic diagram of the proposed VLC architecture using COTS hardware components.
As shown in the figures, the proposed VLC transceiver design contains the following components: host PCs, WiFi NICs, local oscillators (LO), RF mixers, LED driver circuits, LED, photodetector and proper types of cables and connectors.
Here the RF signal emitted by the WiFi NIC of each antenna port is down-converted to meet the specification of the VLC front-end.
An up-conversion happens for received signal at VLC front-end so that it can be injected into the WiFi receiver.
Note that the up-/down-conversion is required as COTS WiFi chipsets integrate baseband processing unit and radio transceiver in a single system-on-chip (SoC) and expose only RF signal in 2.4\,GHz or 5\,GHz band.

\subsection{Hardware Setup}\label{section:hardware}
\begin{figure}[ht]
    \centering
    \includegraphics[width=0.75\linewidth]{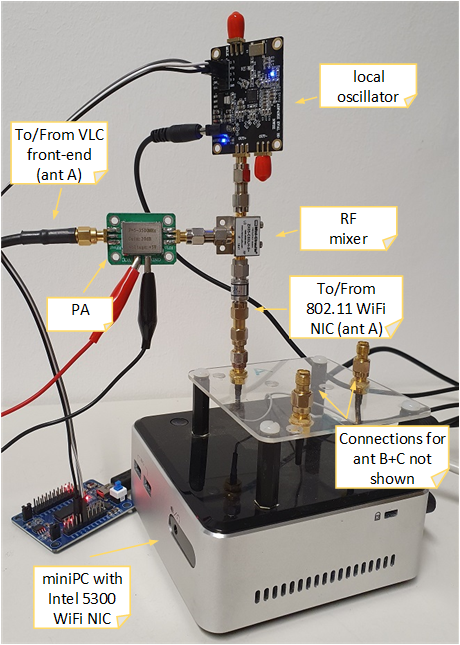}
    \caption{VLC prototype using COTS hardware.}
    \label{fig:tx_prototype}
    \vspace{-5pt}
\end{figure}

Fig.~\ref{fig:tx_prototype} shows a photo of our VLC transceiver prototype.
The specification, functionality and price of each hardware component is provided in Table~\ref{vlc_hardware}.
The total cost of the hardware providing one TX and one RX chain is around 130\$ for a single node. Note that the cost of the host PC and VLC front-ends is not included.

\begin{table}[ht!]
\centering
\captionsetup{justification=centering}
\caption{The list of hardware components used for the implementation of a VLC SISO node (prices from 2020).}
\label{vlc_hardware}
\footnotesize
\begin{tabular}{c|c|c|c}
Type     & Name                                                                 & Description                & Price          \\ \hline

WiFi NIC & \begin{tabular}[c]{@{}c@{}} Intel 5300 \\ (or any other)\end{tabular} & 
\begin{tabular}[c]{@{}c@{}}Generation and processing \\ of WiFi RF signal \end{tabular} & $\approx$ \$15 \\  \hline

\begin{tabular}[c]{@{}c@{}}15\,dB \\ Attenuator\end{tabular} & No Name & \begin{tabular}[c]{@{}c@{}}To match RF mixer's \\ voltage input requirements\end{tabular} & $\approx$ \$10 \\  \hline

2 $\times$ RF Mixer & \begin{tabular}[c]{@{}c@{}}Mini-Circuits \\ ZX05-C60-S+\end{tabular} & \begin{tabular}[c]{@{}c@{}}Up-/down-conversion \\ of WiFi RF signal \end{tabular}  & $\approx$ \$35 \\ \hline

VLO & ADF4351 & LO signal for RF mixer & $\approx$ \$15 \\ \hline

USB controller & CY7C68013A & For control of VLO & $\approx$ \$10 \\ \hline

\begin{tabular}[c]{@{}c@{}}Power \\ Amplifier\end{tabular}
  & No Name & \begin{tabular}[c]{@{}c@{}}For power amplification \\ of TX signal\end{tabular} & $\approx$ \$10 \\ 

\end{tabular}
\end{table}

A WoV node consists of a host PCs, e.g. a small form factor PC like Intel NUC, equipped with a single COTS WiFi network interface cards (NIC), that generates and consumes 802.11-compliant waveform.
On the software side, we do not modify any components and use the standard Linux operating system and proper driver modules matching used NICs.
In particular, our solution is transparent to upper layers, which allows us to reuse entire existing protocol stacks and run any application on top of VLC communication link, e.g. \texttt{iperf}.

\subsection{Base-band Signal}
We use the base-band signals from unmodified WiFi NICs and down/up converted to match the input/output of the optical front-ends.

\noindent \textbf{Tx Side:} The wireless NIC (in our case Intel Wi-Fi Link 5300) generates an analog radio frequency (RF) signal (i.e. waveform of WiFi frame), providing three (one for each antenna port) single-ended I (in-phase) and Q (quadrature) outputs. 
The RF signal parameters (i.e. center frequency and bandwidth) are determined by the pre-configured 802.11 WiFi channel, while the usage of multiple antennas depends on selected modulation and coding scheme (MCS), i.e. number of spacial streams. 
For example, channel 1 has a bandwidth of 20\,MHz and is centered at 2412\,MHz, while MCS values between 0-7 employ only one spatial stream over single antenna. 
Note that Intel wireless driver allows selecting the operating antennas (in both directions, i.e. TX and RX) by exposing proper registers in its \texttt{debugfs} directory. This way, we set one antenna port to operate as TX chain and second port as RX chain. 
Our VLC front-end (see next section) modulates the LED with a any signal between 0-100\,MHz, hence, we down-convert the WiFi signal accordingly.
To this end, the single-ended I/Q output of wireless NIC is connected to the single-ended RF input of RF mixer.
The controllable ADF4351 oscillator is connected to single-ended LO input of RF mixer and provides a signal with 5\,dBm power. 
The oscillator requires an external 5\,V power supply that can be obtained from the USB port using a USB to jack power cable adapter.
The TX power of Intel NIC is too high for used RF mixer.
Therefore, in order to avoid non-linear behavior and match the operating range, we weaken the signal with a 15\,dB attenuator.
In order to shift the WiFi signal to the required frequency range, we set the oscillator frequency to 2375\,MHz.
This way, the WiFi signal at channel 1 (i.e. 2412\,MHz) is down-converted to the center frequency of 2412-2375 = 37\,MHz.
To control the frequency of VLO, we use the FX2LP CY7C68013A USB micro-controller together with open-source software \texttt{pyadf435x}\footnote{https://github.com/jhol/pyadf435x}.
The down-converted signal is available at the IF output of the RF mixer. 
However, due to the attenuation and down-conversion (conversion loss of around 5.6\,dB) the signal is too weak to drive the LED, hence, we have to amplify it by at least +5\,dB using power amplifier (PA). The PA requires 5V power supply, that again can be obtained from USB port. 
In the evaluation section, we show that amplification by +20\,dB greatly extends the communication distance, however, it also distorts the signal effectively preventing usage of higher-order modulations.
Finally, the signal is input to the driver circuit, where it is used to modulate the intensity of the LED. 
Note that the signal is effectively up-converted from MHz to THz band.

\noindent \textbf{Rx Side:} The setup on the receiver side is an inverse connection of the transmitter side. 
The photo-detector (PD) outputs the signal between 0-100\,MHz. 
As shown in Fig.~\ref{fig:wifi_over_vlc_arch}, the output of PD is connected to the RF mixer, which up-converts the WiFi signal from 37\,MHz to 2412\,MHz. 
Note that we connect the second output port of the VLO into RF mixer's LO port.
The signal can be up-converted to any valid WiFi channel and we use this possibility in our evaluation in order to avoid any possible cross-talks over RF channels, i.e. the transmitter operates at channel 1 while the receiver at channel 6. Our experiments reveal that the expected RF cross-talks cannot be received by collocated nodes, therefore the usage of two RF channels is not needed.
The signal is then sent to WiFi NIC though one of available antenna ports. 
The NIC performs baseband signal processing to recover the signal and decode the original data frame. Note that NIC itself is equipped with an automatic gain control (AGC) module, therefore we do not amplify the signal.

\subsection{VLC Front-ends}
The VLC transmitter hardware consists of an LED driver and a broadband optical front-end using red light-emitting diodes (LED)~\cite{HamamatsuPD}. 
In addition, lenses of different field-of-view (FOV) may be attached to the optical front-end. 
The driver modulates the incoming voltage signal into the instantaneous optical power of the LED. 
As the optical power can be modulated between zero and maximal value, the input signal cannot be negative and a proper biasing is required. 
To this end, the driver circuit adds an incoming AC signal to the DC bias. 
In order to support transmissions with higher-order MCS, the LED driver provides linear operation in a wide input signal range.
This feature is especially important in case of OFDM as it is subject to high peak-to-average power ratios.

The VLC receiver hardware consist of highly sensitive, broadband front-end with concentrators glued onto the receiver photo-diodes (PD)~\cite{CreeLED}. 
The PD translates the light intensity into the photo-current, which is converted into a voltage signal by a built-in linear Transimpedance Amplifier (TIA).

\begin{figure}[ht]
  \vspace{-5pt}
  \centering
  \hfill
  \begin{minipage}[ht!]{0.45\linewidth}
    \includegraphics[width=1.0\linewidth]{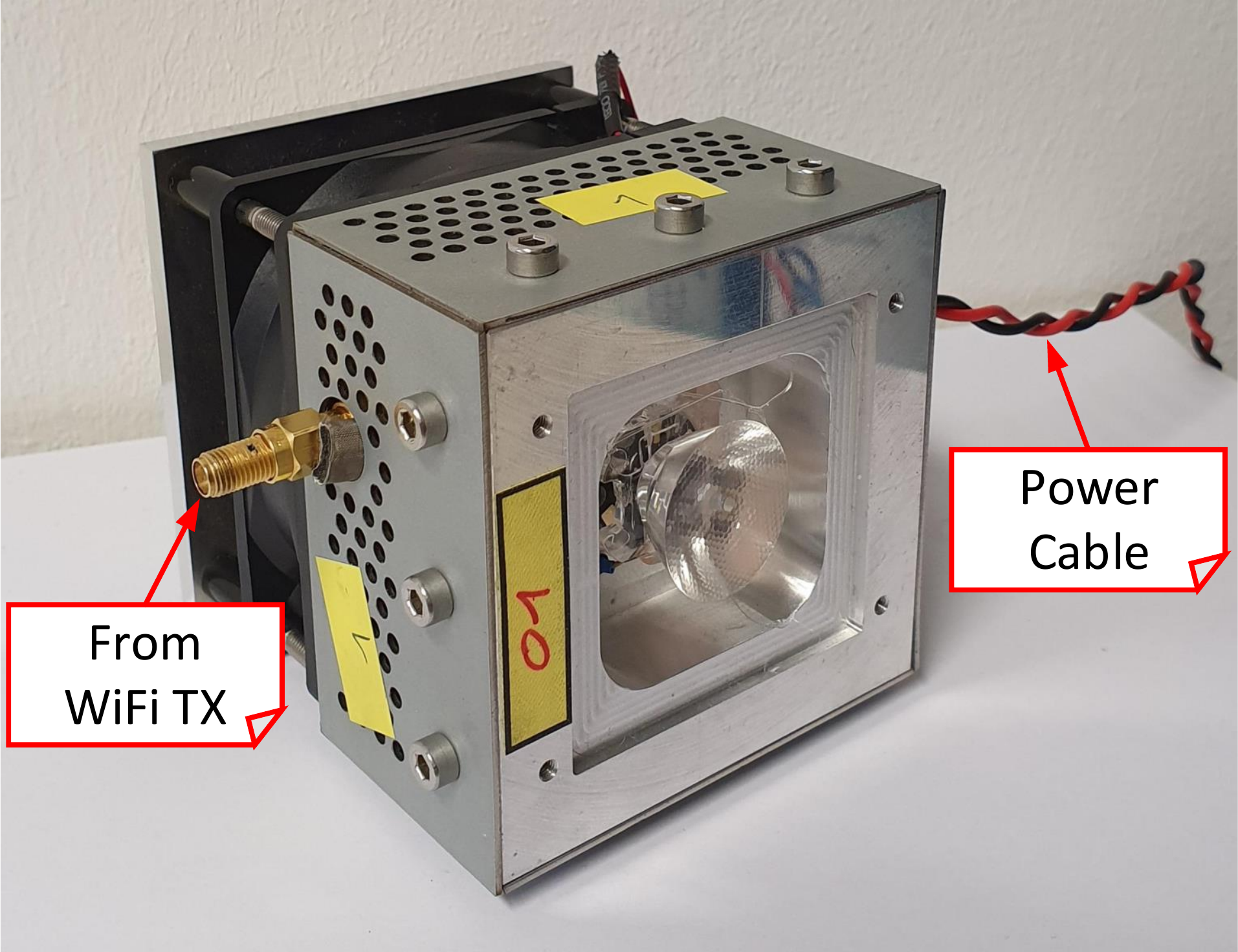}
  \end{minipage}\hfill
  \hfill
  \begin{minipage}[ht!]{0.45\linewidth}
    \includegraphics[width=1.0\linewidth]{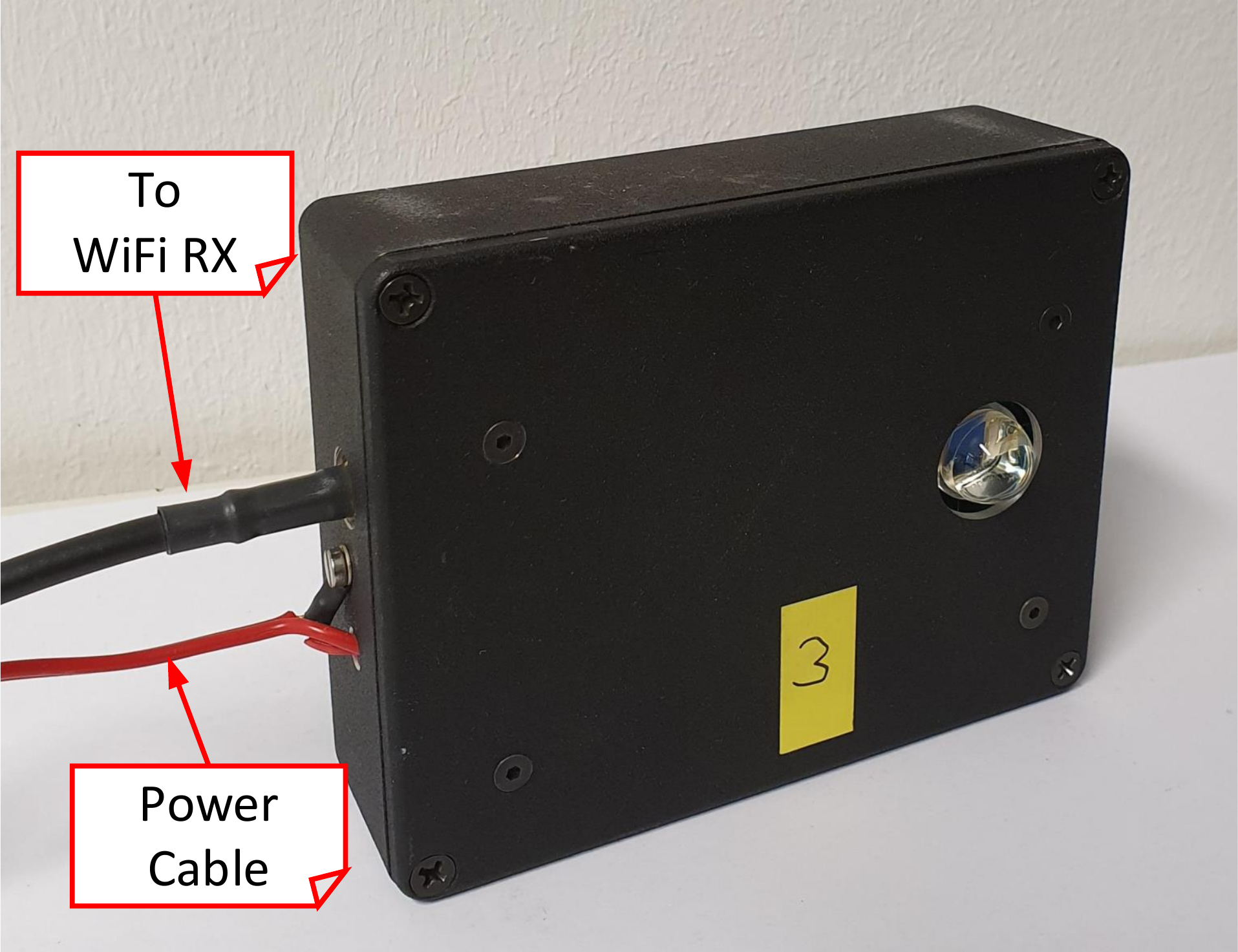}
  \end{minipage}\hfill
  \caption{VLC front-ends: Tx (left) and Rx (right).}
  \label{fig:vlc_frontends}
  \vspace{-5pt}
\end{figure}

The VLC front-ends are designed and developed by Fraunhofer HHI in Berlin -- Fig.~\ref{fig:vlc_frontends}.

\section{Evaluation}
In this section we present results from experiments using our prototype.
We analyze the performance of a single unidirectional VLC link.
First, we investigate the additional distortions introduced due to the chosen architecture.
Second, we analyze the relationship between frame error rate and receive signal strength.
Third, we analyze the maximum communication distance.
\subsection{Methodology}
Our prototype is analyzed by means of experiments in a small testbed.
The hardware and software configuration used for the WoV nodes was already described in section~\ref{section:hardware}.
\subsection{Results}

\begin{figure}[ht!]
    \centering
    \includegraphics[width=1.0\linewidth]{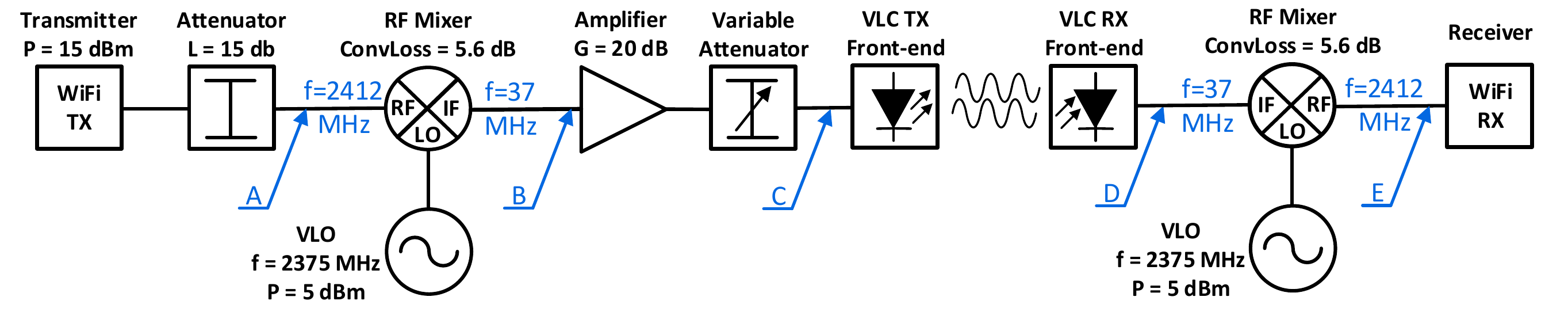}
    \vspace{-5pt}
    \caption{Tx/Rx chain with probing points A-E shown.}
    \label{fig:txrx_chain}
    \vspace{-5pt}
\end{figure}

\noindent\textbf{Experiment 1: (Distortions)}
The objective is to check the correctness of the WiFi signal fed into a VLC front-end by analyzing the distortions introduced in the TX/RX pipeline due to the usage of RF mixers and attenuators.
Therefore, we recorded the signal with a Software-defined Radio (SDR) platform (Ettus USRP X310 with UBX-160 Daughterboard) at different stages of the TX/RX processing pipeline (Fig.~\ref{fig:txrx_chain}) and analyzed its low-level statistics and the constellation diagrams using Matlab WLAN Toolbox~\cite{matlabWlan}.
Moreover, for the different stages of the pipeline, we estimated the received power at each OFDM subcarrier.

\medskip

\begin{figure*}[t]
  \vspace{0pt}
  \begin{minipage}[b]{0.196\linewidth}
    \caption*{Probe point A}
    \vspace{-5pt}
	\includegraphics[width=\linewidth]{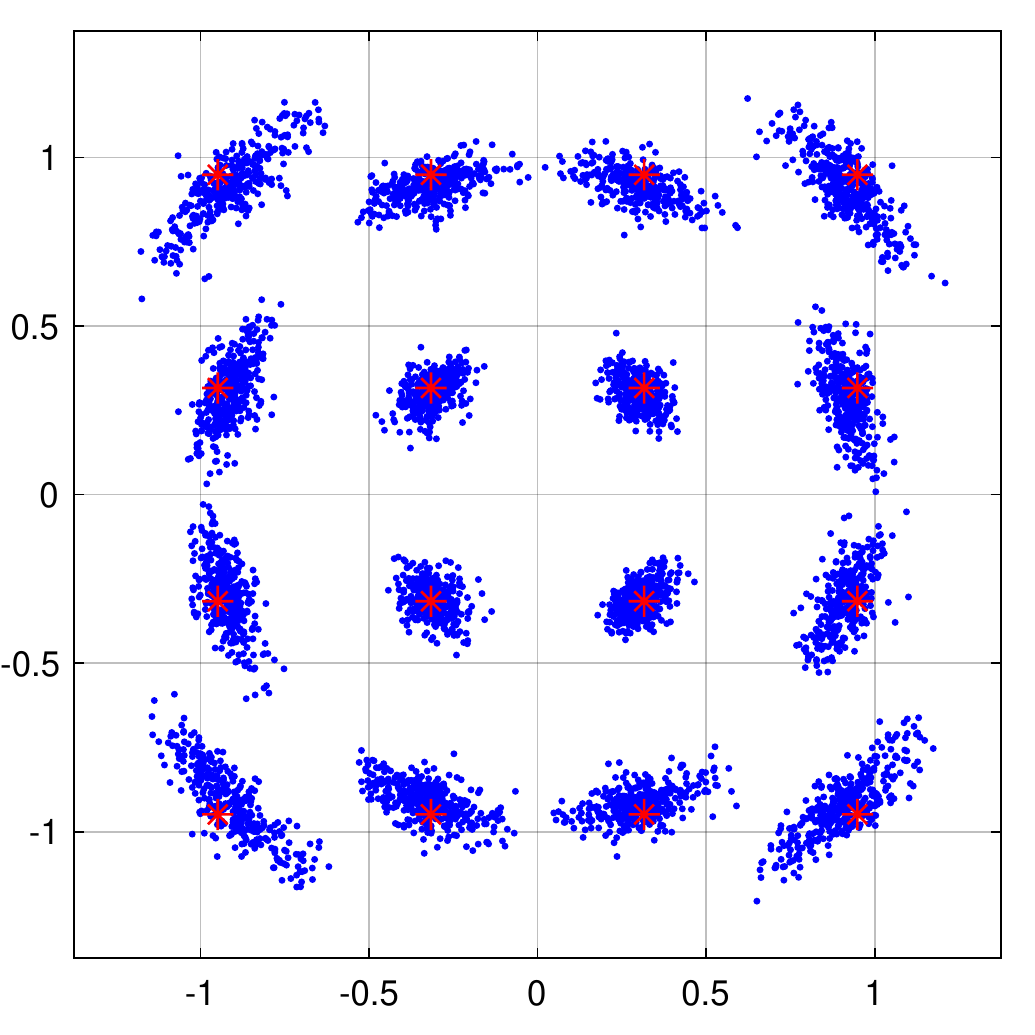}
    \vspace{-5pt}
    \label{fig:constDiagram_A}
    \vspace{-5pt}
  \end{minipage}\hfill
  \begin{minipage}[b]{0.196\linewidth}
    \caption*{Probe point B}
    \vspace{-5pt}
	\includegraphics[width=\linewidth]{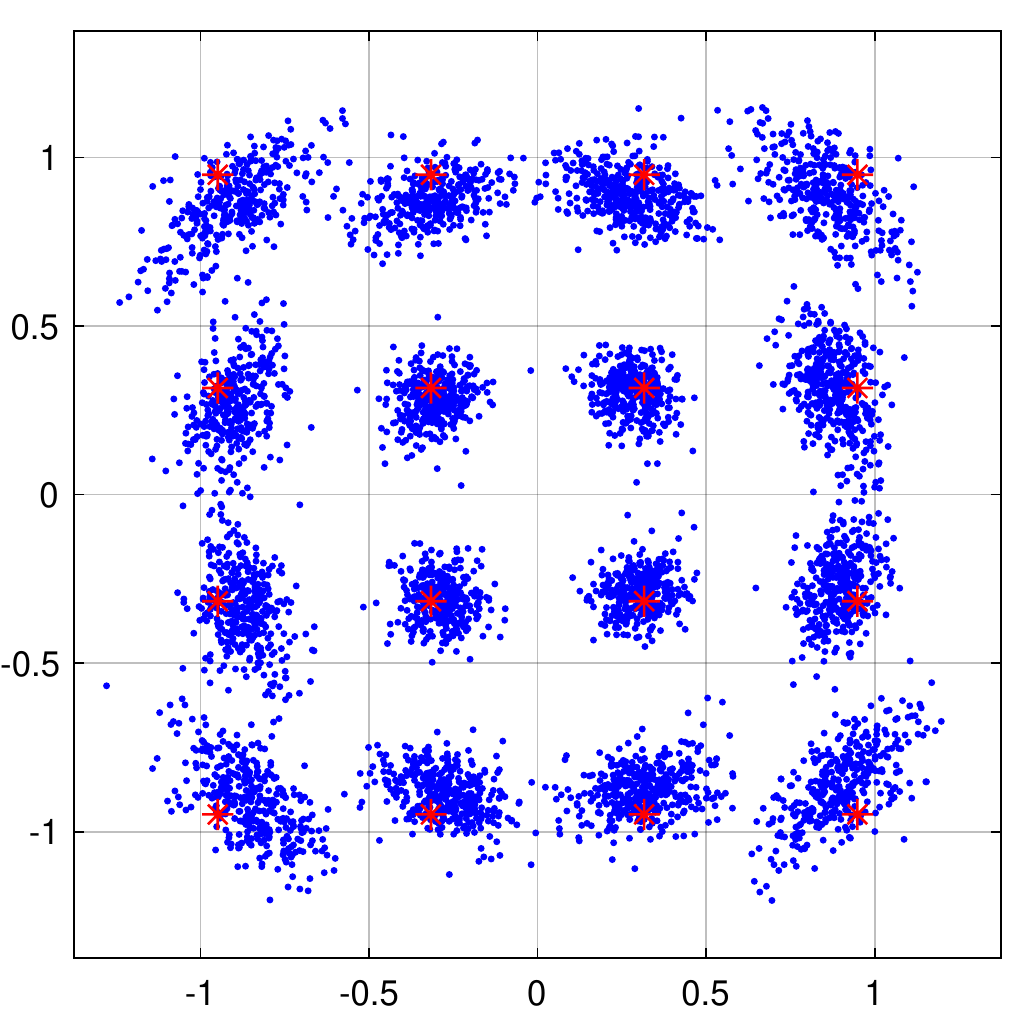}
    \vspace{-5pt}
    \label{fig:constDiagram_B}
    \vspace{-5pt}
  \end{minipage}\hfill
  \begin{minipage}[b]{0.196\linewidth}
    \caption*{Probe point D}
    \vspace{-5pt}
	\includegraphics[width=\linewidth]{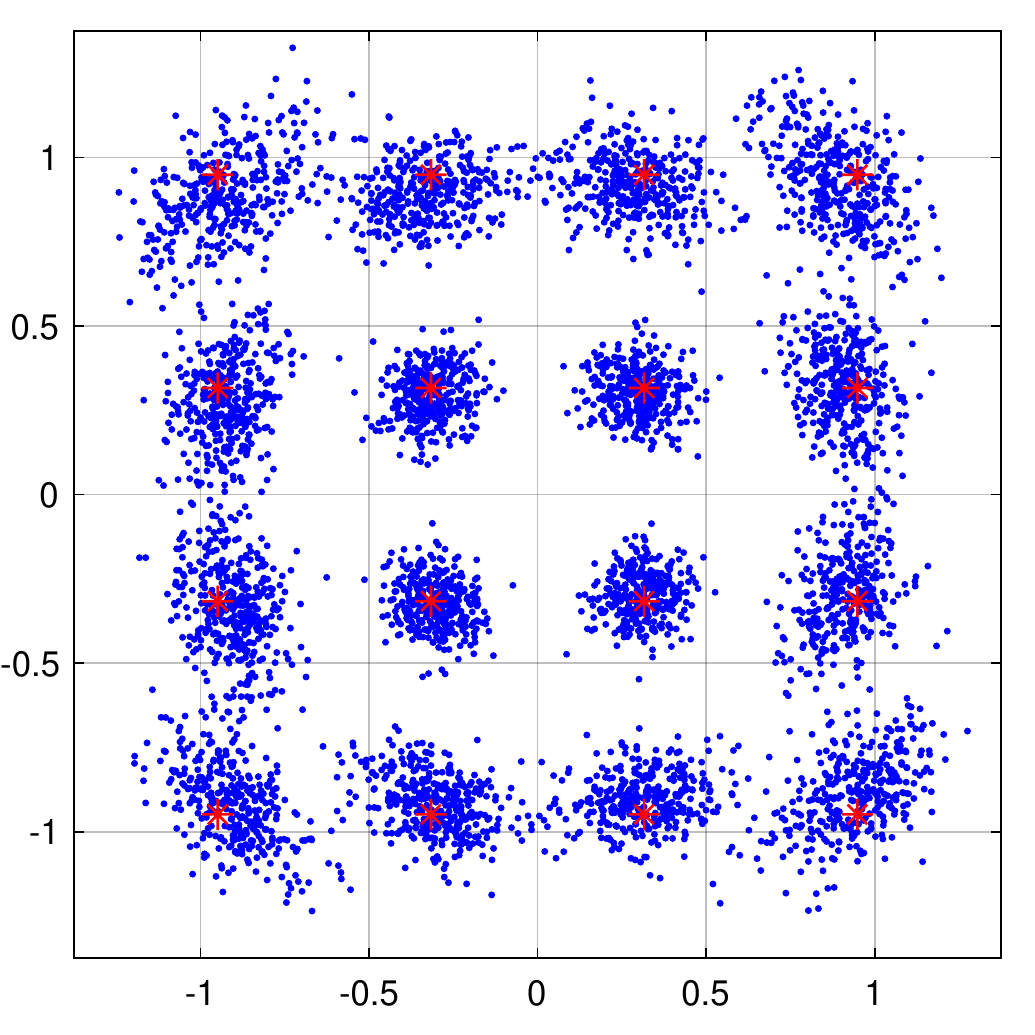}
    \vspace{-5pt}
    \label{fig:constDiagram_D}
    \vspace{-5pt}
  \end{minipage}\hfill
  \begin{minipage}[b]{0.196\linewidth}
    \caption*{Probe point E}
    \vspace{-5pt}
	\includegraphics[width=\linewidth]{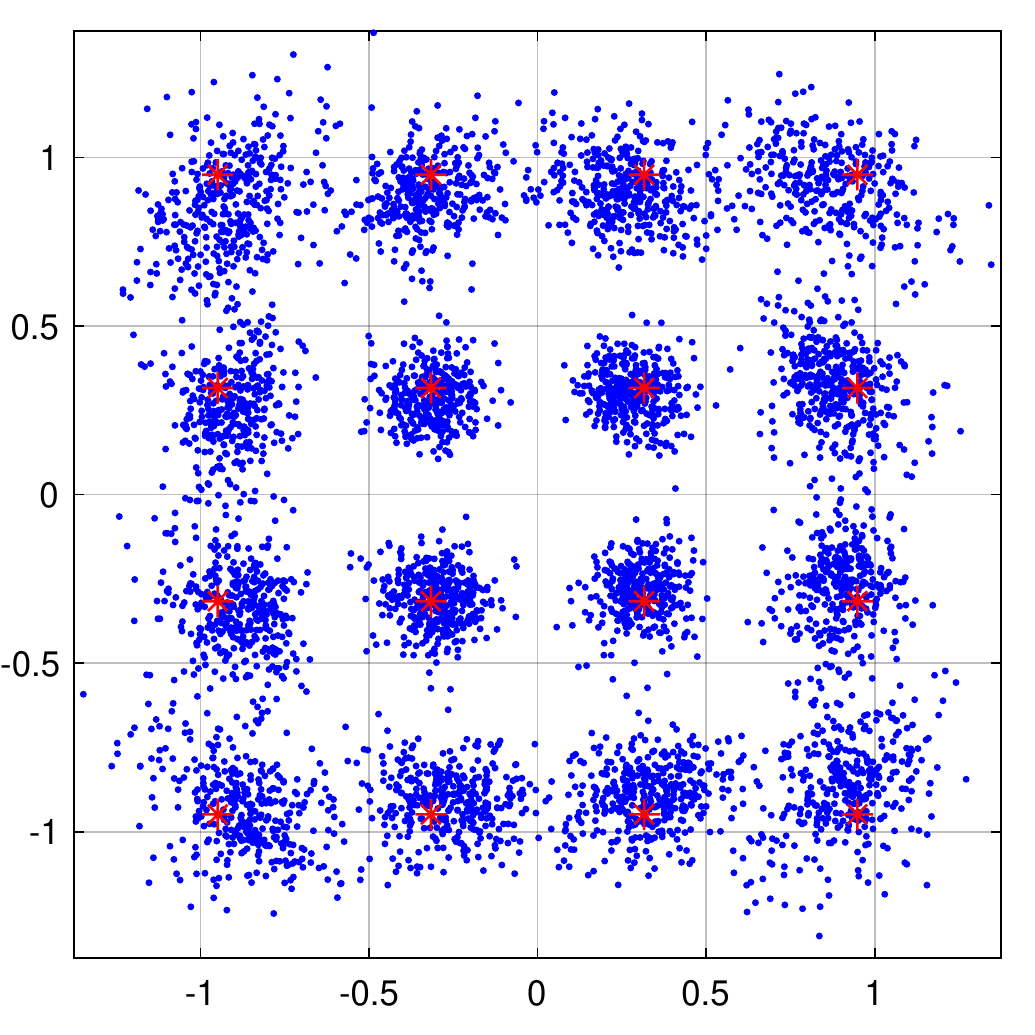}
    \vspace{-5pt}
    \label{fig:constDiagram_E}
    \vspace{-5pt}
  \end{minipage}\hfill
  \begin{minipage}[b]{0.204\linewidth}
    \caption*{Receive power}
    \vspace{-8pt}
	\includegraphics[width=\linewidth]{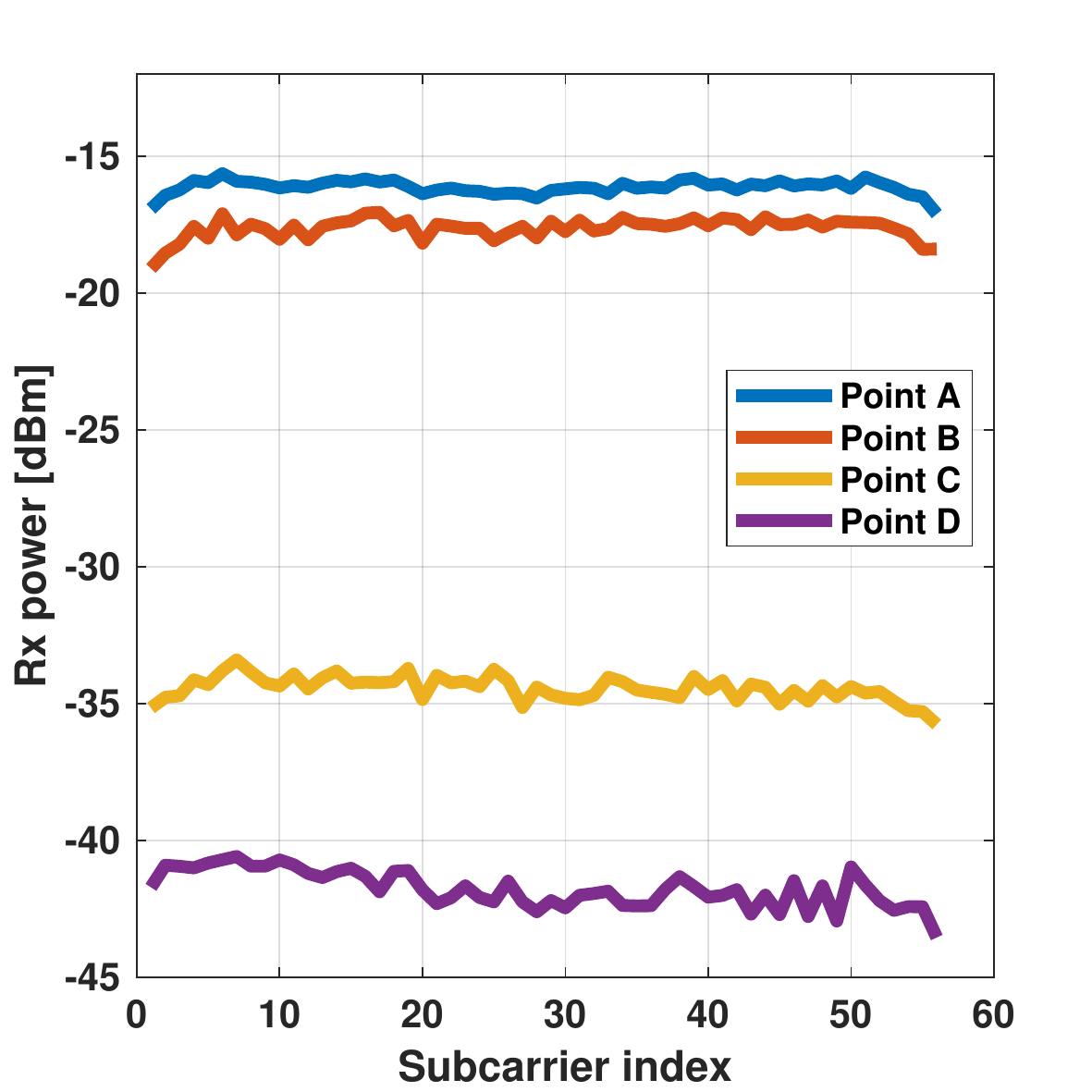}
    \vspace{-5pt}
    \label{fig:constDiagram_A}
    \vspace{-5pt}
  \end{minipage}\hfill
  \vspace{-5pt}
  \caption{QAM constellation diagrams and receive power at each OFDM subcarrier obtained from CSI at selected probe points.}
  \vspace{-10pt}
  \label{fig:constallation_diagrams}
\end{figure*}

\noindent\textbf{Result 1:}
Fig.~\ref{fig:constallation_diagrams} shows the measured constellation diagrams of 16-QAM transmission (MCS 3) at the different stages. We can see that even the signal coming directly from Intel WIFI NIC (i.e. Point A) suffers from a high phase noise level. At point B, the signal loses around 2\,dB of power due to conversion loss, which is lower than expected 5.6\,dB as specified for the used RF mixer. We believe that it is caused by very high power of the signal fed into the RF port, that is not the case in normal operation. We were not able to collect the signal at point C as its power was too high for the USRP platform (i.e. clipping effect). At point D and E, again, the signal's power is lower due to path-loss over VLC link (distance of 50\,cm) and conversion loss, respectively. In Table~\ref{table:matlab}, we show low-level signal statistics as obtained using 802.11n receiver developed in Matlab WLAN Toolbox.
The received power at each OFDM subcarrier confirms that the VLC channel is frequency-flat.
Our results prove that the proposed approach is valid as up-/down-conversion does not distort the WiFi signal, i.e. it introduces only a small loss of around 1\,dB in SNR.

\begin{table}[ht!]
\centering
\footnotesize
\caption{Low-level statistics of the WiFi signal at different probe points obtained using Matlab WLAN Toolbox.}
\label{table:matlab}
\begin{tabular}{c|c|c|c|c}
Check Point & \begin{tabular}[c]{@{}c@{}}Rx Power Level\\ {[}dBm{]}\end{tabular} & \begin{tabular}[c]{@{}c@{}}Noise Level\\ {[}dBm{]}\end{tabular} & \begin{tabular}[c]{@{}c@{}}SNR\\ {[}dB{]}\end{tabular} & \begin{tabular}[c]{@{}c@{}}EVM\\ {[}\%{]}\end{tabular} \\ \hline
A &   1.18  & -29.53 & 32.32 & 11.70 \\
B &  -0.56  & -32.31 & 31.14 & 12.47 \\
C &    ---  & ---    & ---   & ---   \\
D &  -17.14 & -40.67 & 23.52 & 15.35\\
E &  -24.53 & -45.85 & 22.04 & 16.22\\
\end{tabular}
\vspace{-5pt}
\end{table}

\noindent\textbf{Experiment 2: (SISO Link Performance)}
The objective is to analyze the performance of a single unidirectional WoV link.
Specifically, we investigate the relationship between receive signal strength (RSSI) as reported by the WiFi NIC and the achieved Frame Success Rate (FSR) for Modulation and Coding Schemes (MCS) between 0 (BPSK, 1/2) and 7 (64QAM, 3/4).
Moreover, we analyze the impact of channel bonding available in 802.11n, i.e. 20\,MHz vs. 40\,MHz channel.
Therefore, the experiment is composed of a WoV transmitter and receiver node, each with a single antenna port connected to VLC front-end, placed at different distances from each other.
The unicast 802.11n frames with different MCS with acknowledgment disabled are injected using the WiFi NIC in monitor mode.
For each measurement point, we transmitted 1000 frames of the size of 1000\,Bytes and checked the number of received frames with correct CRC checksum.

\medskip

\begin{figure}[ht]
    \centering
    \includegraphics[width=0.85\linewidth]{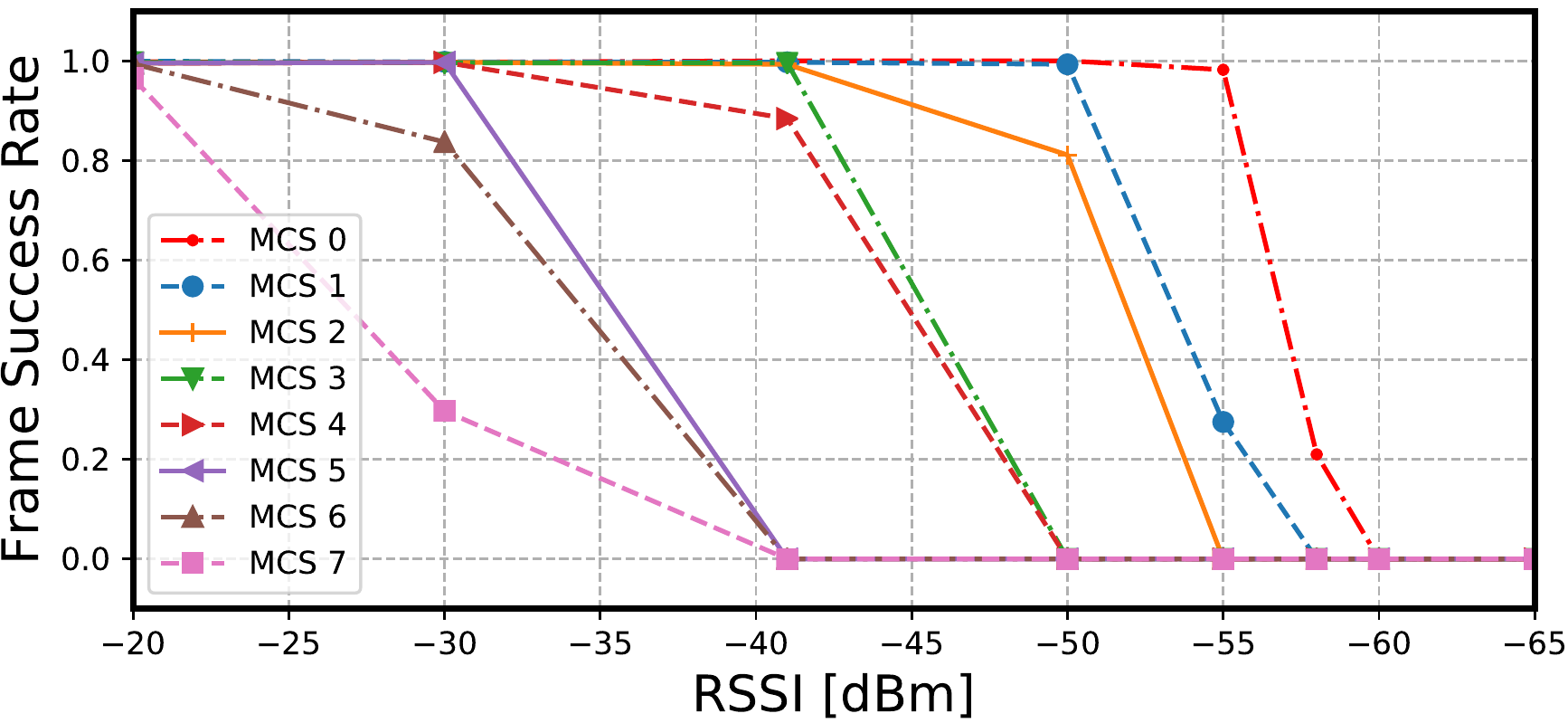}
    \vspace{-5pt}
    \caption{Frame Success Rate vs. RSSI (20\,MHz channel).}
    \label{fig:fsr_sisio_20m}
    \vspace{0pt}
\end{figure}

\noindent\textbf{Result 2:}
Fig.~\ref{fig:fsr_sisio_20m} shows the relationship between RSSI and FSR for all MCSs defined in the 802.11n standard. 
We can see that for the MCS 0 (i.e. BPSK 1/2) an RSSI of -55\,dBm is sufficient to have the FSR of almost 1.0.
For the MCS 7 the signal must be around 30\,dB stronger to have same FSR.
From those results and the known SNR thresholds for different 802.11n MCS~\cite{halperin2011predictable} we can determine the noise floor of our WoV setup to be at around -60\,dBm.

\begin{figure}[ht]
    \centering
    \includegraphics[width=0.85\linewidth]{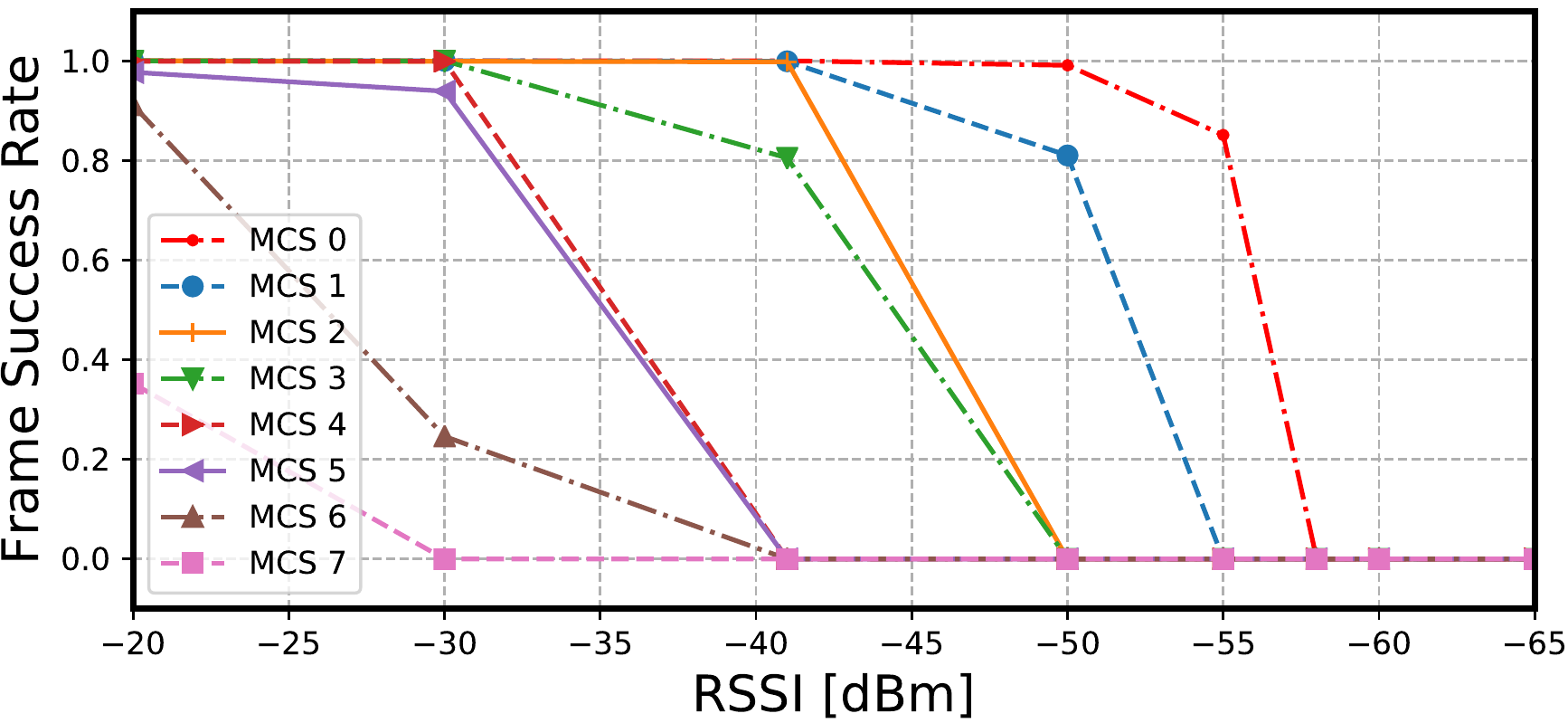}
    \vspace{-5pt}
    \caption{Frame Success Rate vs. RSSI (40\,MHz channel).}
    \label{fig:fsr_sisio_40m}
    \vspace{-5pt}
\end{figure}

Fig.~\ref{fig:fsr_sisio_40m} shows the performance in case of channel bonding.
The results are similar to 20\,MHz channel except that the curves are shifted by around 3\,dB what caused by the increased noise floor level~\cite{halperin2011predictable}.
At the highest RSSI which corresponds to a distance of around 25\,cm a physical layer bitrate of 150\,Mbit/s (i.e. MCS 7 with short guard interval) is possible.

\noindent\textbf{Experiment 3: (Max Communication Distance)}
The goal of this experiment is to study the maximum communication distance of our VLC prototype and investigate the impact of usage of a signal power amplifier (PA) and optical lenses.

\medskip

\begin{figure}[ht]
    \centering
    \vspace{-5pt}
    \includegraphics[width=0.85\linewidth]{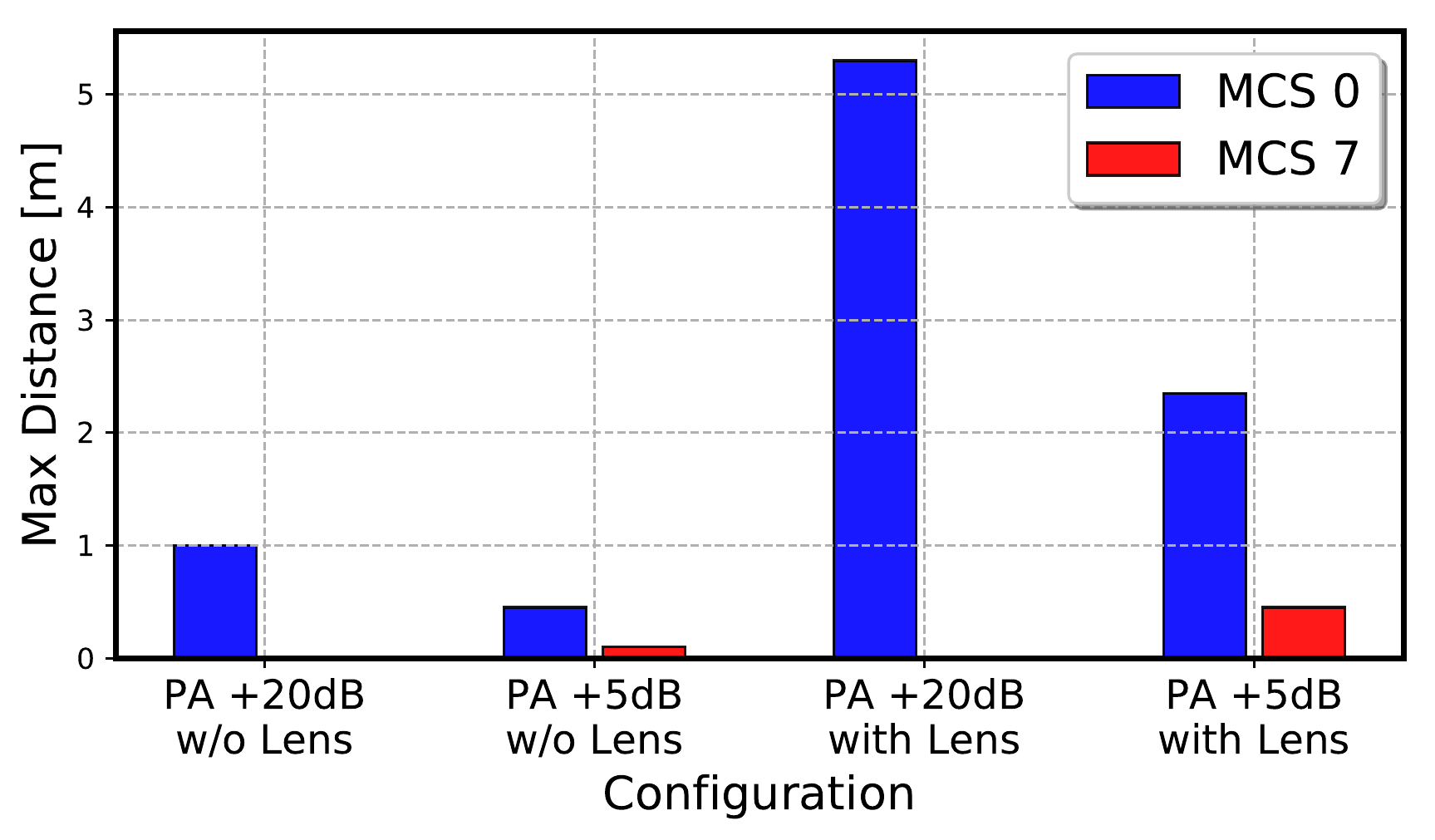}
    \vspace{-5pt}
    \caption{Communication distances for different setups.}
    \label{fig:max_dist_sisio}
    \vspace{-10pt}
\end{figure}

\noindent\textbf{Result 3:}
Fig.~\ref{fig:max_dist_sisio} shows the maximum achievable communication distance under different configurations and 20\,MHz channel.
Using a PA with 20\,dB together with a lens at the VLC front-end transmitter a distance of up to 5\,m is possible using MCS 0 (i.e. data rate of 7.2\,Mbit/s). However, amplification by 20\,dB causes non-linear distortions to the signal (i.e. clipping effect) in our TX-RX chain and prevents communication using higher-order MCSs that use QAM.

%
\section{Discussion}

While the integration of COTS WiFi chipsets with VLC front-ends provides satisfactory results allowing for real-time communication it suffers from the following limitations:

\noindent \textbf{Bandwidth:} In our VLC testbed, we rely on signal processing capabilities of COTS WiFi hardware. 
As a result, the maximum bandwidth of our testbed is limited by the bandwidth of the WiFi channel. 
Although the newest WiFi standards allow to use channel bonding with up to 160\,MHz, it is still only a fraction of the visible light spectrum.

\noindent \textbf{Frequency Range:} The Mini-Circuits RF mixer can up-/down-convert the signal between 0-6000\,MHz. 
However, the ADF4351 oscillator generates signal only up to 4400\,MHz with a step of 25\,MHz. 
Therefore, in the current version of our WoV testbed, we are limited to use only WiFi signals from 2.4\,GHz band where maximal channel bandwidth is 40\,MHz.
Note that the usage of different signal generators (e.g. Maxim Integrated MAX2870 that operates up to 6000\,MHz)ł enables down-conversion of WiFi signals from 5\,GHz band and usage of up to 160\,MHz channels.

\noindent \textbf{Waveform:} In our VLC testbed, we rely on the waveforms supported by 802.11 standard and implemented by WiFi NIC.
Currently it is classical OFDM and DSSS from legancy 802.11b standard.
There is no easy way use custom waveforms specifically designed for VLC like optical OFDM~\cite{shieh2008coherent}.

Despite these limitations, we believe that our easy-to-use WoV testbed is expected to have many applications in prototyping VLC transceivers ranging from link level measurements to system-level evaluations.

%
\section{Related Work}

\noindent\textbf{OOK-based VLC platforms:}
Schmid et al.~\cite{Schmid} presented a VLC system based on light bulbs equipped with a simple System-on-a-Chip (SoC) running an embedded version of Linux. 
In order to generate the optical channel, the micro-controller modulates the light intensity using Pulse Width Modulation (PWM) based on simple \textit{on-off keying} (OOK). 
The maximal data rate equals 700\,b/s. 
The authors provided a transparent integration of VLC communication channel with Linux networking stacks.
Specifically, they developed a driver module that expose network interface towards Linux networking stack and encodes incoming data frames into PWM signal.
OpenVLC~\cite{Galisteo2} is a simple, low-cost hardware/software platform meant for prototyping software-based and programmable MAC and PHY protocols for VLC. 
It is centered around a printed circuit board (i.e. OpenVLC cape) that implements an optical front-end. 
The cape is attached to BeagleBone Black single-board computers running Linux operating system. 
The authors provided OpenVLC Linux driver module that implements the MAC and PHY layers. 
The software module provides simple primitives (including sampling, symbol detection, coding/decoding, channel contention and carrier sensing) that a researcher can use to build her own VLC MAC and PHY protocols. 
Moreover, the driver module provides integration with TCP/IP protocol stack enabling Internet connectivity. 
Using simple \textit{on-off keying} modulation, the newest version of the platform (i.e. OpenVLC 1.3) achieves data rate of up to 400\,kb/s.
In contrast, our approach is completely transparent and requires no changes nor development of custom driver modules as we simply reuse the existing well-developed and tested WiFi subsystem and its integration into Linux OS.
Moreover, since we transmit ordinary WiFi signals, we can access very detailed information about physical properties of underlying communication channel, e.g. channel state information.
Finally, this makes our system suitable for the development of VLC solutions for broadband services.

\noindent\textbf{OFDM-based VLC platforms:}
Amjad et al.~\cite{amjad2019ieee} proposed an experimental vehicular VLC system, which integrates a custom-made driver hardware, commercial vehicle light modules, and an open-source implementation of 802.11a physical layer in GNURadio.
In contrast to our approach it is only suitable for link-level studies as no real MAC layer exists.
Moreover, the physical layer is very limited and outdated, i.e. it covers just basic functions of 802.11a/p and misses advanced WiFi features like usage of multiple antennas from newer standard like 802.11n/ac/ax.
Our approach evolves as the WiFi NICs can easily be replaced by newer generations of 802.11 like WiFi 6 (802.11ax).
%

%
\section{Conclusions}
While our first prototype demonstrates the feasibility of WiFi over VLC using COTS devices, we plan to extend our work to support bi-directional VLC communication as well as advanced physical layer schemes of 802.11 like the use of multiple-input and multiple-output (MIMO) systems.
The use of receive antenna diversity techniques (e.g., maximal-ratio combining) for the application in VLC is very promising as it is very efficient when dealing with mobility induced handover and shadowing related issues like signal blockage which is very common in VLC environments.
Moreover, using MIMO together with wide channels (up-to 160\,MHz in 802.11ac/ax) the achievable data rate can be dramatically increased reaching multiple Gbps.
Finally, we plan to design and develop printed circuit board, that will integrate all of the used components in a compact way, eliminate the RF connectors (hence, the losses introduced by them) as well as minimize per-unit costs which is of importance when building large MIMO systems.

\section*{Acknowledgement}
We are grateful to Fraunhofer HHI for providing us the VLC front-ends.
This work was supported by the German BMBF under grant agreement No. 16KIS0985 (OTB-5G+ project).

\bibliographystyle{IEEEtran}
\bibliography{biblio,IEEEabrv}

\end{document}